\documentclass[]{spie}
\usepackage[dvips]{graphicx}

\title{Measurements of Si Hybrid CMOS X-Ray Detector Characteristics} 

\author{Stephen D. Bongiorno \supit{a}, Abe D. Falcone \supit{a}, David N. Burrows \supit{a}, Robert Cook \supit{a}, Yibin Bai \supit{b}, Mark Farris \supit{b}
\skiplinehalf
\supit{a}Pennsylvania State University, Dept. of Astronomy \& Astrophysics, University Park, PA 16802 USA; \\
\supit{b}Teledyne Imaging Sensors, 5212 Verdugo Way, Camarillo, CA 93012 USA
}

\authorinfo{Further author information: (Send correspondence to S.D.B.)\\S.D.B: E-mail: sdb210@astro.psu.edu, Telephone: 1 814 865 8485}

\begin{document}
\maketitle

\begin{abstract}
The development of Hybrid CMOS Detectors (HCDs) for X-Ray telescope focal planes will place them in contention with CCDs on future satellite missions due to their faster frame rates, flexible readout scenarios, lower power consumption, and inherent radiation hardness. CCDs have been used with great success on the current generation of X-Ray telescopes (e.g. Chandra, XMM, Suzaku, and Swift).  However their bucket-brigade readout architecture, which transfers charge across the chip with discrete component readout electronics, results in clockrate limited readout speeds that cause pileup (saturation) of bright sources and an inherent susceptibility to radiation induced displacement damage that limits mission lifetime.  In contrast, HCDs read pixels with low power, on-chip multiplexer electronics in a random access fashion.  Faster frame rates achieved with multi-output readout design will allow the next generation's larger effective area telescopes to observe bright sources free of pileup.  Radiation damaged lattice sites effect a single pixel instead of an entire row.  Random access, multi-output readout will allow for novel readout modes such as simultaneous bright-source-fast/whole-chip-slow readout.  In order for HCDs to be useful as X-Ray detectors, they must show noise and energy resolution performance similar to CCDs while retaining advantages inherent to HCDs.  We will report on readnoise, conversion gain, and energy resolution measurements of an X-Ray enhanced Teledyne HAWAII-1RG (H1RG) HCD and describe techniques of H1RG data reduction.
\end{abstract}


\section{INTRODUCTION\label{introduction.sec}}
CCDs are operating as X-ray imaging spectrographs at the focal planes of nearly all major imaging X-ray telescopes currently in orbit.  The technology is in a high state of maturity and it performs very well in terms of readnoise and non-dispersive energy resolution.  However, future focusing X-ray observatories such as the International X-ray Observatory (IXO), Xenia, and Gen-X will look very different from this generation's workhorse observatories.  Future missions plan to use $>10 \times$ larger effective areas to accomplish their science goals and, given similar plate scales, this corresponds to an order of magnitude increase in flux density at the focal plane.  CCDs would not perform well at the focal planes of such telescopes due primarily to saturation (pileup).  In collaboration with Teledyne Imaging Sensors, we are developing novel X-ray Hybrid CMOS designed to satisfy the hardware requirements of next generation X-ray observatories.

The advantages of hybrid CMOS over CCD technology are due to the HCD's flexible readout architecture; that is, their ability to address and non-destructively read single pixels via the read out integrated circuit (ROIC) instead of transferring charge across the entire chip raster-style as with CCDs.  The generation of HCDs currently being tested have 16 outputs, meaning 16 pixels can be addressed at a given instant.  Multiple outputs, combined with the ability to directly address any pixel, decreases pileup by allowing small windowed regions around bright sources to read out at a fast frame rate, while the remaining pixels are read out at a slow frame rate.  The advantage comes from achieving high frame rates on bright sources without increasing the pixel rate and subsequently increasing readnoise.  An additional advantage to reading out the array with the ROIC is that CMOS process technology is inherently low power compared to CCD readout, an important consideration on space missions.

The design differences between CCDs and HCDs also give the hybrid CMOS chip inherent radiation hardness advantages over CCDs.  Proton and alpha particle radiation will damage a detector when particles enter the substrate and displace silicon nuclei.  This disruption of the otherwise periodic crystal lattice locally alters the band gap structure, potentially creating unwanted potential wells or ``traps'' within the bulk lattice.  Charge carriers passing through a trap will be held in the trap.  In a CCD, the charge in pixels upstream of a damaged pixel is forced to pass through the trap.  This degrades the efficiency with which charge is moved from pixel to pixel (charge transfer efficiency (CTE)) in the entire column, smearing charge opposite the read direction.  In an HCD, however, charge only travels through the $\sim 100~{\rm \mu m}$ thickness of the absorber layer as opposed to the $\sim 2~{\rm cm}$ across it.  Proton displacement damage will affect a single pixel as opposed to the majority of a column, making HCDs significantly more radiation hard than CCDs.

Finally, HCDs are also inherently more resistant to micrometeoroid damage than CCDs.  The gate and line structures of a front illuminated (FI) CCD or CMOS sensor are relatively close to the substrate surface and therefore vulnerable to micrometeoroid impact.  With the critical electronics located below the absorber layer, in the mux, an HCD will be more resistant to impact damage.  As with radiation hardness, micrometeoroid damage will likely be contained to a small number of pixels in HCDs, similar to the damage observed on XMM-Newton's back-illuminated pn-CCD.\cite{struder01}  This is in contrast to the catastrophic damage observed on Suzaku's XIS2 front illuminated CCD in November 2006, which is thought to have been the result of micrometeoroid impact.


\section{HYBRID CMOS DETECTORS}
Hybrid CMOS detectors are radically different from their CCD counterparts.  Both technologies use biased silicon semiconductor as a medium for X-ray photon absorption, photo-charge generation, and charge collection.  The term {\it hybrid} in {\it hybrid CMOS detector} refers to the fact that each detector is composed of two substrates that are individually optimized for their purpose and then bonded together.  In the HyViSI H1RG, the absorber substrate is a silicon PIN photo-diode array; its only purpose is to absorb X-rays and convert their energy into electron-hole pairs.  The bottom layer, covered by the absorber layer and therefore not illuminated, is a silicon substrate upon which the readout integrated circuit (ROIC) logic is built for each pixel.  The layer is referred to as the multiplexer, or mux, because the electronics enable digitally choosing an input source (a single pixel) and routing that signal to a detector output trace.  The multiplexer is what makes CMOS detectors unique because each pixel can be addressed individually, routed to one of multiple outputs, and non-destructively read or reset.

Hybrid CMOS technology has been well established by Teledyne for use in the infrared/optical\cite{bai04}, with a 2048$\times$2048 HgCdTe version of the H2RG planned to fly on the James Webb Space Telescope.  Accordingly, the detector was named H1RG for HgCdTe Astronomy Wide Area Infrared Imager 1024$\times$1024 with Reference pixels and Guide mode, referencing the chip's IR/optical origins.  This detector development project, which is a collaboration between Penn State University and Teledyne Imaging Sensors, is taking advantage of the large amount of work that went into developing flight quality detector and readout hardware for JWST and adapting it for X-ray use.


\section{PSU TEST STAND} 
This paper will report results from characterizing H1RG-125, an 18 $\mu$m pixel pitch detector with a 500 \AA\ aluminum optical blocking filter deposited on half of the frontside surface.  This detector was manufactured in 2006.  Our initial work concentrated on testing the new HyViSI H1RG chips with discrete component readout hardware, establishing readnoise and energy resolution measurements.  However, the most recent work, which is reported here, concentrates on testing the H1RG using a SIDECAR\texttrademark controller.  The SIDECAR\texttrademark connects to the H1RG package through a 92 line flex cable and  supplies clocks and biases, and performs chip programming, signal amplification, analog to digital conversion, and data buffering.  Both the H1RG-125 and the SIDECAR\texttrademark are mounted inside a light-tight vacuum chamber, with the detector fastened to a coldfinger typically held at 150K and the SIDECAR\texttrademark mounted to the room temperature chamber wall.  A shuttered $^{55}$Fe source is positioned in front of the detector.  

\section{THE DATASET \& TEST OPTIMIZATION}
In this characterization report, we analyze a dataset consisting of 97 psuedo correlated double sample (CDS) images with the array exposed to an $^{55}$Fe source.  In a typical CCD, the CDS is an amplifier baseline voltage subtraction performed in hardware during the readout of each pixel.  H1RG-125 does not contain on-chip CDS, so we perform a similar function in software that we call pseudo CDS.  The pseudo CDS consists of two frames, one where the array is reset, immediately read, and written to file row by row, and a second where the array is simply read and written to file.  In software, the first frame is subtracted from the second to produce the pseudo CDS frame. 

Each CDS image contains approximately 3600 events and the entire dataset contains 352950 events.  The data was acquired in 100 kHz unbuffered, two output mode, meaning that two, 512 pixel wide columns are each read out simultaneously at the rate of 100 pixels per second.  The preamp gain was optimized at 18.06 dB = 8$\times$ to take full advantage of the ADC's dynamic range.  To match Teledyne's test conditions, the detector was maintained nominal at $T=150$K.  In an effort to reduce charge spreading, the substrate bias voltage was set to $V_{sub} = 18$V.  Given the 100 kHz pixel rate, 1024$\times$1024 array size, and 2 channel output, the resulting integration time is 5.24 seconds. 

\subsection{FET LEAKAGE CURRENT}
An issue that was encountered while first attempting to use the SIDECAR\texttrademark to read out the H1RG was the observation of a severe vertical gradient in each output channel of the image that did not subtract out during CDS subtraction.  This was discovered to be due to leakage current in the SIDECAR\texttrademark's preamp FET's reference voltage.  To solve the problem we worked with Teledyne to create a custom readout scheme where the FET is reset at the end of each row.  Teledyne engineers informed us that this leakage current follows the same temperature dependence of the Shockley diode equation, where $I_L\propto e^{(1/T)}$, and is only significant when the SIDECAR\texttrademark is run at room temperature.  Teledyne predicts that the leakage current would be negligible if the SIDECAR\texttrademark were to be cooled, however this test has yet to be performed.

\subsection{ROW NOISE CORRECTION\label{rownoise.sec}}
\begin{figure}
\centering
\includegraphics[width=0.5\textwidth]{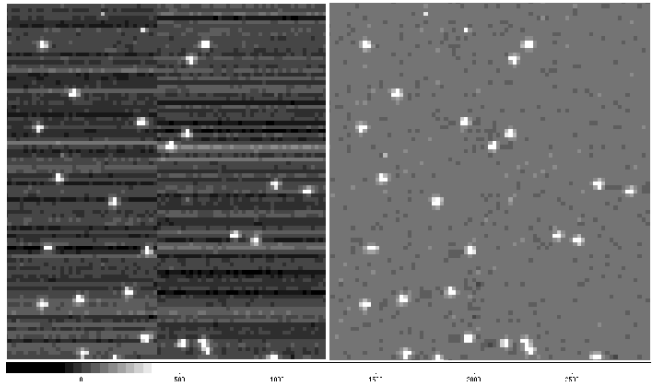}
\caption{A CDS subtracted $^{55}$Fe data image before ({\it left}) and after ({\it right}) median filter subtraction to reduce row noise.  Note the flat background in the median filter processed image.  Bright dots are X-ray events.}
\label{med_sub.fig}
\end{figure}

Following the pseudo CDS subtraction, we are still left with horizontal artifacts, or ``row noise", in the images.  We first attempted to remove the row noise by making use of the H1RG's reference pixels, a four pixel wide border of mux pixels surrounding the chip which are connected to ordinary capacitors rather than a PIN photo-diode on the substrate array.  The reference pixels are designed to enable the removal of constant amplifier offset.  However, the row noise that we observed had nonlinear structure, requiring more effort than subtracting a single number off of each row.  Therefore, subtracting off reference pixel values did not work well.  To solve the problem we subtract the result of a 15 pixel moving median filter off of each row, treating each detector output channel individually.  This technique was successful, producing images free of horizontal artifacts.  Figure \ref{med_sub.fig} shows an image before and after median filter subtraction.

\section{INTERPIXEL CAPACITANCE}
\begin{figure}
\begin{center}
\includegraphics[width=0.5\textwidth]{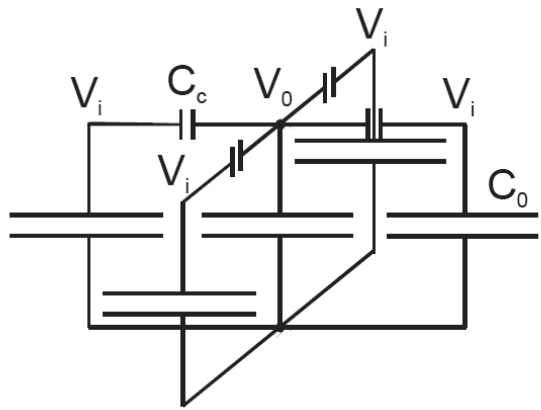}
\end{center}
\caption{\label{capcup_diagram.fig}An electronics schematic of a pixel and its top, bottom, left, and right neighbors\cite{finger06}.  There is an expected capacitance between the pixel and readout node, $C_0$, and also a parasitic capacitance between adjacent pixels, $C_c$.  This IPC causes integrated charge to spread between neighboring pixels.}
\end{figure}

\begin{figure}
\begin{center}
\includegraphics[width=0.5\textwidth]{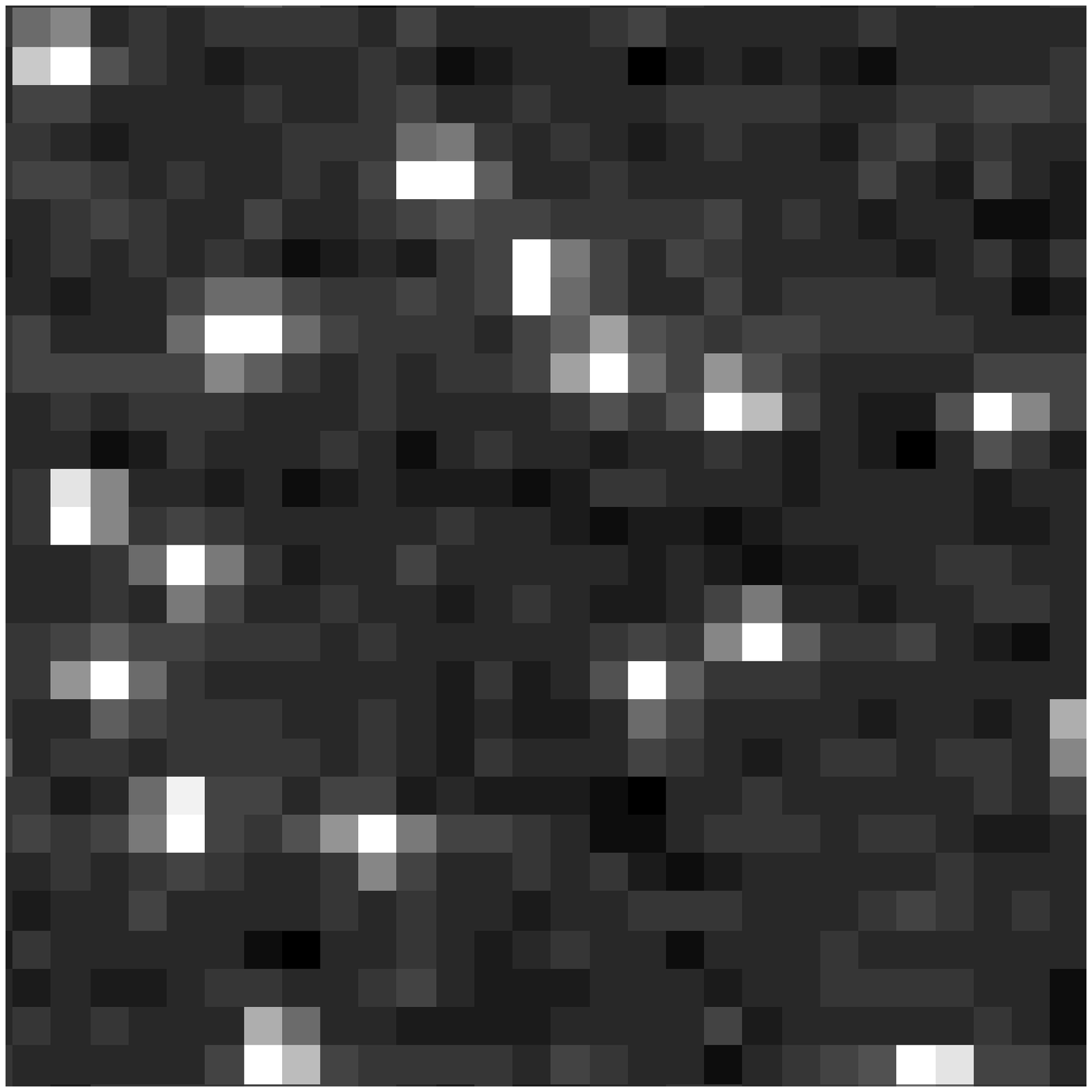}
\end{center}
\caption{\label{capcup_data.fig}A 28x28 image sub-array that shows charge spreading effects in $^{55}$Fe X-ray events.  In a CCD, X-ray events are typically confined to one or two pixels.  However, in HyViSI H1RG arrays, events typically spread to 8 or 9 pixels.}
\end{figure}

Interpixel Capacitance (IPC) is a physical effect in HyViSI HxRG detectors that gives rise to image artifacts.\cite{finger06}  The effect is caused by unintended, parasitic capacitances between adjacent pixels inside the frontside detector substrate.  Figure \ref{capcup_diagram.fig} illustrates a simplified circuit diagram that shows both the capacitance between each pixel and its readout node, $C_0$, and the IPC, $C_c$.  The resulting cross shaped spreading of the detector point spread function (PSF) can be seen in the small sub-array from an X-ray data image shown in Figure \ref{capcup_data.fig}.

X-rays are convenient for characterizing the IPC because very often they will deposit charge in one or two pixels, effectively acting as an impulse function.  To characterize the PSF we look only at X-ray events where the eight pixels surrounding the brightest, center pixel have a standard deviation less than or equal to the 1$\sigma$ noise floor of the image.  The intent of the analysis is to pick out the most symmetrical events that originally contained all of their charge in one pixel before IPC took effect.  The resulting PSF is shown in Table \ref{capcup_result.tab}.  The IPC of HyViSI H1RG detectors has been characterized by Finger et al.\cite{finger06}, though their method is different and involves producing an inverse IPC pattern by continually resetting a 1 pixel window region in an array under uniform illumination.  Their result, truncated into a normalized 3$\times$3 region to match this work, is shown in Table \ref{capcup_result.tab}.

\begin{table}
	\begin{center}
		\begin{tabular}{cc}
			\begin{tabular}{|c|c|c|}
				\multicolumn{3}{c}{Averaging} \\
				\hline
				0.024 & 0.072 & 0.027 \\
				\hline
				0.077 & 0.591 & 0.081 \\
				\hline
				0.026 & 0.073 & 0.024 \\
				\hline
			\end{tabular}
			\begin{tabular}{|c|c|c|}
				\multicolumn{3}{c}{Window Reset} \\
				\hline
				0.0222 & 0.0603 & 0.0218 \\
				\hline
				0.0698 & 0.6681 & 0.0583 \\
				\hline
				0.0211 & 0.0567 & 0.0212 \\
				\hline
			\end{tabular}
		\end{tabular}
	\end{center}
	\caption{The 3$\times$3 IPC PSF obtained by averaging symmetrical events in our dataset ({\it left}) and the IPC PSF of a different HiViSi H1RG using a 	1x1 pixel window reset method\cite{finger06}({\it right}).}
	\label{capcup_result.tab}
\end{table}

\section{GRADING}
As we have shown, when an X-ray interacts with an HCD, the signal is spread over multiple pixels.  Grading is the process of identifying regions of the array where single X-rays have generated photocharge in the substrate and recording properties of the event.  We are particularly interested in which pixels surrounding the interaction site contain signal, and also the morphology of that pixel arrangement.  

Our first step in event grading is to identify events.  We require that a pixel satisfy a primary threshold value that is set $\approx4\sigma$ above the noise floor and also be a local maximum, greater than its 8 surrounding pixels, in order to be considered an event.  Next, we identify which of the 8 pixels in the surrounding 3$\times$3 region contain significant signal by requiring that they satisfy a $3 \sigma$ secondary event threshold in order to be counted in the event sum.  Setting the secondary threshold as high as possible allows fewer surrounding pixels to be counted towards the sum and minimizes the readnoise to enter the event.  However, setting the secondary threshold too high causes signal to be excluded from the event sum.  These parameters must be tuned.

In CCD X-ray instruments, events are also assigned a number according to their morphology.  This grade number indicates which pixels in the event satisfy the secondary threshold and are used to separate good events from bad detections.  However, due to the $\approx40$\% charge spreading effect of IPC in our HCD, the direct application of morphology grades has not proved useful in analysis.  Instead, to quantify event morphology, we calculate the fraction of the primary event pixel contained in the second brightest pixel; simply the value of the second brightest pixel in an event divided by the value of the brightest pixel.  We call this quantity the ``percent split''.

\begin{figure}
\centering
\includegraphics[width=0.5\textwidth]{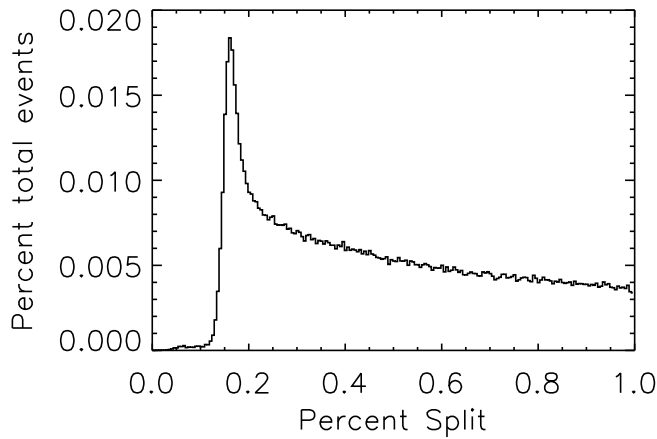}
\caption{The percent split is calculated by dividing the value of the second brightest pixel in an event by the value of the brightest.  This quantity provides a simple measure of how much spreading the event exhibits.  Plotting a histogram of the percent split values for every event in the dataset shows that a greater relative fraction of events have a percent split of $\sim 0.2$.  We believe that these were single pixel events prior to IPC.}
\label{psplit_hist.fig}
\end{figure}

To visualize what the percent split of all events looks like, we plot a histogram of the parameter in Figure \ref{psplit_hist.fig}.  The strong low percent split peak visible in the histogram indicates that a larger than average fraction of events exhibit a percent split between 0.1 and 0.2.  Compared to other events in the dataset, these low percent split events have a larger fraction of their signal contained within the center pixel.  This leads us to believe that these were single pixel events prior to IPC, and will therefore be the best events to use in energy resolution characterization.

\section{SYSTEM GAIN}
The gain of the entire X-ray detector system is the number, in units of digital number (DN) per electron volt (eV), that represents the complete conversion of incident photon energy into a digital number.  The total system gain can be broken down into several factors:
\begin{eqnarray*}
G (DN/eV) = \frac{1}{3.65}~\left(\frac{e^-}{eV}\right)~\frac{1}{Capacitance}\left(\frac{Volts}{e^-}\right)~Preamp~Gain\left(\frac{ADC~DN}{ADC~Range~(Volts)}\right)\\
\end{eqnarray*}
In silicon, on average, one electron is promoted from the valence to conduction band for every 3.65 eV of energy in the absorbed X-ray.  When the charge in a pixel is read, the capacitance of the output node determines the voltage signal produced by a given amount of charge in the pixel.  Finally this voltage is amplified and converted into a digital number at an ADC.  We measure the total system gain by producing graded event spectra and observing the DN value of the $^{55}$Mn 5.9 K$_{\alpha}$ keV, 1613.69 e$^-$ X-ray line.  

\begin{figure}
\centering
\includegraphics[width=0.5\textwidth]{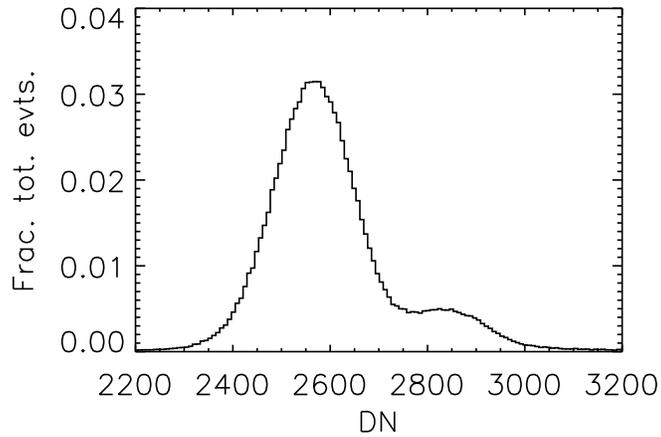}
\caption{Varying the secondary event threshold during event grading will impact the energy spectra.  Here, a low 3$\sigma$ event threshold was used during event grading, resulting in the inclusion of nearly all event island pixels into the event sum.  This increases the line width due to the addition of more readnoise into the sum.}
\label{threesigmaspec.fig}
\end{figure}

\begin{figure}
\centering
\includegraphics[width=0.5\textwidth]{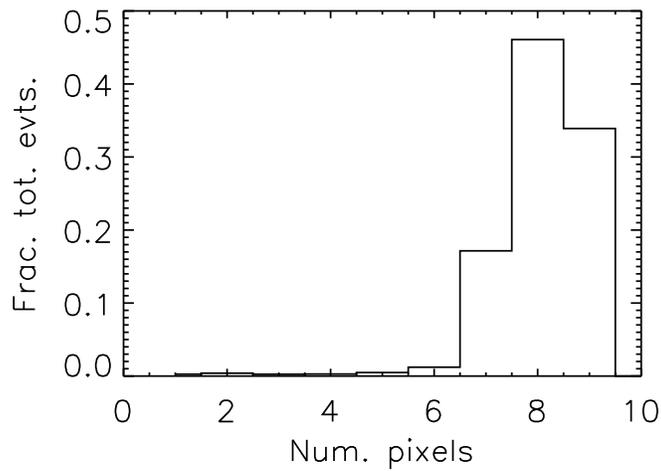}
\caption{Every X-ray that interacts with our detector spreads charge over multiple pixels.  When we want to calculate the system gain, we set the secondary event threshold low (3$\sigma$) to ensure that we catch most of the charge in each event.  This histogram shows the fraction of events in the dataset where some number of pixels satisfied the secondary event threshold.  With a conservatively set secondary threshold, most of the pixels in each event island are included in the sum.}
\label{evt_num_dist.fig}
\end{figure}

Grading is a non-trivial process with significant freedom for parameter tweaking.  To achieve an accurate value for the system gain, we want to ensure that all charge in the pixel region surrounding each event is included in the final spectra.  If some amount of charge were to be lost, then the assumption that every K$_{\alpha}$ photon contains 1613.69 e$^-$ would not be valid.  We accomplish this by using a low, 3 sigma secondary event threshold, and producing an energy spectrum (Figure \ref{threesigmaspec.fig}) consisting almost exclusively of events where all 9 pixels in the 3$\times$3 island are included in the event sum.  This is confirmed by the histogram in Figure \ref{evt_num_dist.fig}, which shows that the largest fraction of all events in the dataset contain 8 pixels that satisfy the given secondary threshold.  The 5.9 keV peak of the 3$\sigma$ secondary threshold energy spectra is located at 2576 DN, making the system gain of 0.626 e$^-$/DN.

\section{NOISE}
We have measured the CDS noise in each frame of our dataset of X-ray images by measuring the FWHM of the noise peak in a histogram computed from image pixel values.  This quantity is not skewed by bright pixels containing X-ray charge and provides a measurement of the noise as data are being taken.  Assuming that the noise is normally distributed allows us to calculate the noise RMS from the pixel histogram FWHM.  Using the previously calculated system gain of 0.626 e$^-$/DN we calculate the CDS noise to be 7.48 e$^-$ RMS.  With no row noise correction we calculate a CDS noise of 27.86 e$^-$ RMS.

In detector test reports, Teledyne calculates this detector's mean CDS noise at 150K to be 12.2 e$^-$ RMS.  Differing array readout, offset correction (such as the row noise correction), noise calculation, or conversion gain may account for the difference.

\section{ENERGY RESOLUTION}
Among the most useful features of X-ray HCDs is that they are imaging spectrographs; they have the ability to do simultaneous imaging and non-dispersive spectroscopy since the energy and location of every X-ray event is recorded by the detector.  The ability of a detector to measure the energy of an incident photon is described by its energy resolution:  $R=\Delta E/E$.  The fundamental limit of energy resolution is set by Fano\cite{fano47}\cite{owens01} noise and a detector that performs at this level is said to be Fano limited.  The noise originates in the uncertainty in the number of electrons generated when an X-ray interacts with a silicon detector.  The variance is:
\begin{eqnarray*}
\sigma^2_F=\frac{F E}{\omega}\\
\end{eqnarray*}
where $F$ is the Fano factor (0.113 for Si), $E$ is the X-ray energy, and $\omega$ is the average electron-hole pair generation energy (3.65 eV for Si).  Inserting numbers, Fano limited performance at 5.898 keV is 0.116 keV, corresponding to an energy resolution of 2\%.  Due to the complicated details of solid state physics that they contain, both $F$ and $\omega$ are always measured empirically.

\begin{figure}
\begin{center}
\includegraphics[width=0.5\textwidth]{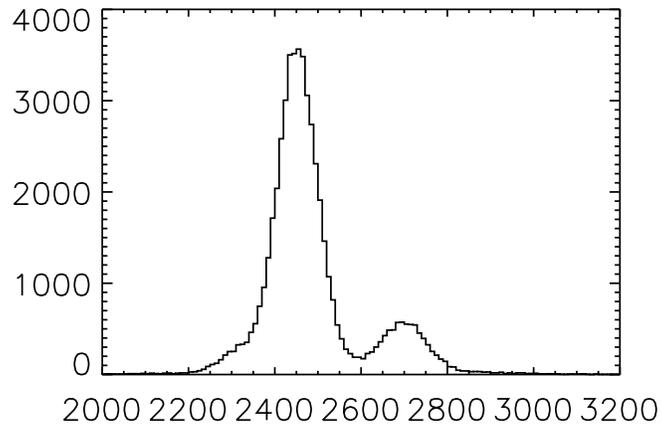}
\end{center}
\caption{\label{spectrum_20split_99th2.fig}An $^{55}$Fe energy spectrum generated by selecting events with percent split less than 0.2, approximately equivalent to selecting single pixel events prior to IPC spreading, and setting the secondary event threshold to 99 DN.  This spectrum has $\Delta E/E$ = 4.2\%.}
\end{figure}

In our system energy resolution is dependent on particular parameters used in data reduction.  The choice of primary and secondary event thresholds and event morphology strongly effect the energy resolution.  Using events with a split percent between 0 and 0.2, and specifying a very large $9\sigma$ secondary event threshold, we obtain the spectrum in Figure \ref{spectrum_20split_99th2.fig}, which has $\Delta E/E$ = 4.2\%.  Specifying such a high event threshold is simply a roundabout method for choosing a particular shape for summing pixels in the events.  95\% of the events were cross shaped after specifying such a high threshold.  Such filtering produces a well resolved spectrum because we require that the events be single or nearly single pixel events before IPC and then effectively choose to sum only the top, bottom, left, right and center pixels.  The method is minimizing the number of pixel reads required to get an accurate measurement of event energy.

\section{SUBSTRATE BIAS VOLTAGE}
\begin{figure}
\centering
\includegraphics[width=0.5\textwidth]{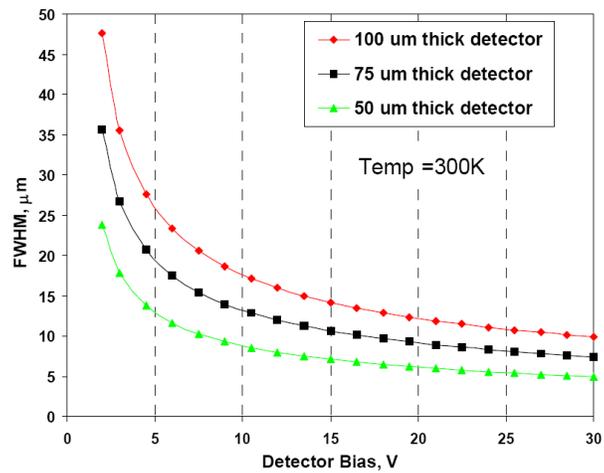}
\caption{\label{vsub_bai.fig}A model of charge cloud spreading as a function of bias voltage in various thickness detectors\cite{bai08}.}
\end{figure}

The bias voltage, $V_{sub}$, across the substrate performs the critical functions of maintaining depletion in the bulk intrinsic silicon and pushing photo-charge clouds into the pixel potential wells.  The higher that $V_{sub}$ is set, the faster a charge cloud will be accelerated into the pixel potential wells.  A faster moving charge cloud has less time for diffusion, which therefore reduces the amount of charge spreading due to charge cloud diffusion.  Teledyne\cite{bai08} has reported this relationship in H1RGs (Figure \ref{vsub_bai.fig}).  While not being able to verify this result completely, we performed an experiment that confirms that a similar relationship exists between $V_{sub}$ and charge spreading.

\begin{figure}
\centering
\includegraphics[width=0.5\textwidth]{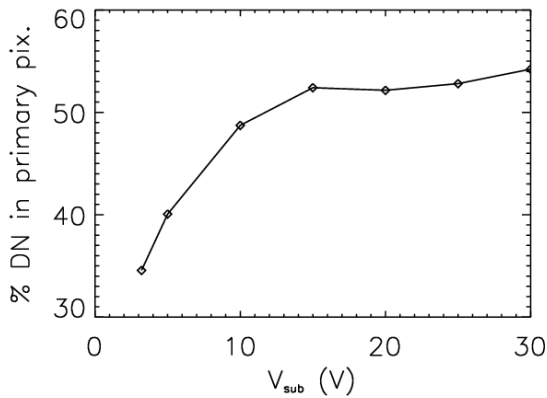}
\caption{\label{vsub_psf.fig}A measurement of percent charge left inside of a local maximum as a function of bias voltage.  Charge spreading is due to both IPC and charge cloud diffusion.}
\end{figure}

To test the relationship, we varied $V_{sub}$ from 3.2 - 30 Volts in 5 Volt increments, acquiring 20,000 events (5 frames) of data at each voltage.  We used 4$\times$4 pixel event regions to ensure that charge spreading farther than 1 pixel distance from the local maximum (primary event pixel) would not be lost.  As a metric for measuring the spread of a given event, we calculated the average percent charge in the primary event pixel as compared to the total charge in the event, for all events at a given voltage.  The results are shown in Figure \ref{vsub_psf.fig}.  In agreement with Teledyne's report that the charge cloud's physical size becomes decreasingly smaller as $V_{sub}$ is increased, we observed that the percentage of charge outside of the primary event pixel decreases with increasing $V_{sub}$.  Our curve appears to abruptly turn over at $\sim$55\%, an effect that we believe is due to a ``saturation" of the measurement whereby IPC prevents charge from being further compressed into the primary event pixel.  In a separate experiment, spatially symmetric events were chosen and their charge spreading percentage measured in order to characterize the inherent PSF of the IPC effect.  It is assumed that charge spreading due to cloud diffusion is minimal in the most symmetric events and that these events correspond to what would be a single pixel event on a CCD.  The results of the experiment were that IPC causes $\sim$59\% of an event's charge to remain in the primary event pixel, which is very close to our observed roll off at $\sim$55\%.  In light of these results, operate our detector with the substrate biased at 18 Volts.

\section{PERMANENT THRESHOLD SHIFT}
It is critical to note that while a reduction in charge spreading from the use of a large $V_{sub}$ may seem like a good idea, operating the H1RG with arbitrarily high voltage across the substrate is potentially dangerous to the H1RG.  Teledyne reported\cite{bai08} that it is possible to damage the CMOS ROIC by oversaturating the array with optical light while substrate is biased at greater than 5.2V.  The damage can occur because, as a pixel becomes deeply saturated, voltage at the pixel's sense node (a source follower gate) will rise towards $V_{sub}$.  CMOS process electronics are designed to withstand $\sim$5.2 Volts.  Therefore, if $V_{sub}$ is set greater than 5.2 Volts and the array allowed to saturate for a large amount of time, the gate oxide will be stressed and the source follower's response can be permanently altered.

\begin{figure}
\begin{tabular}{cc}
\includegraphics[width=0.5\textwidth]{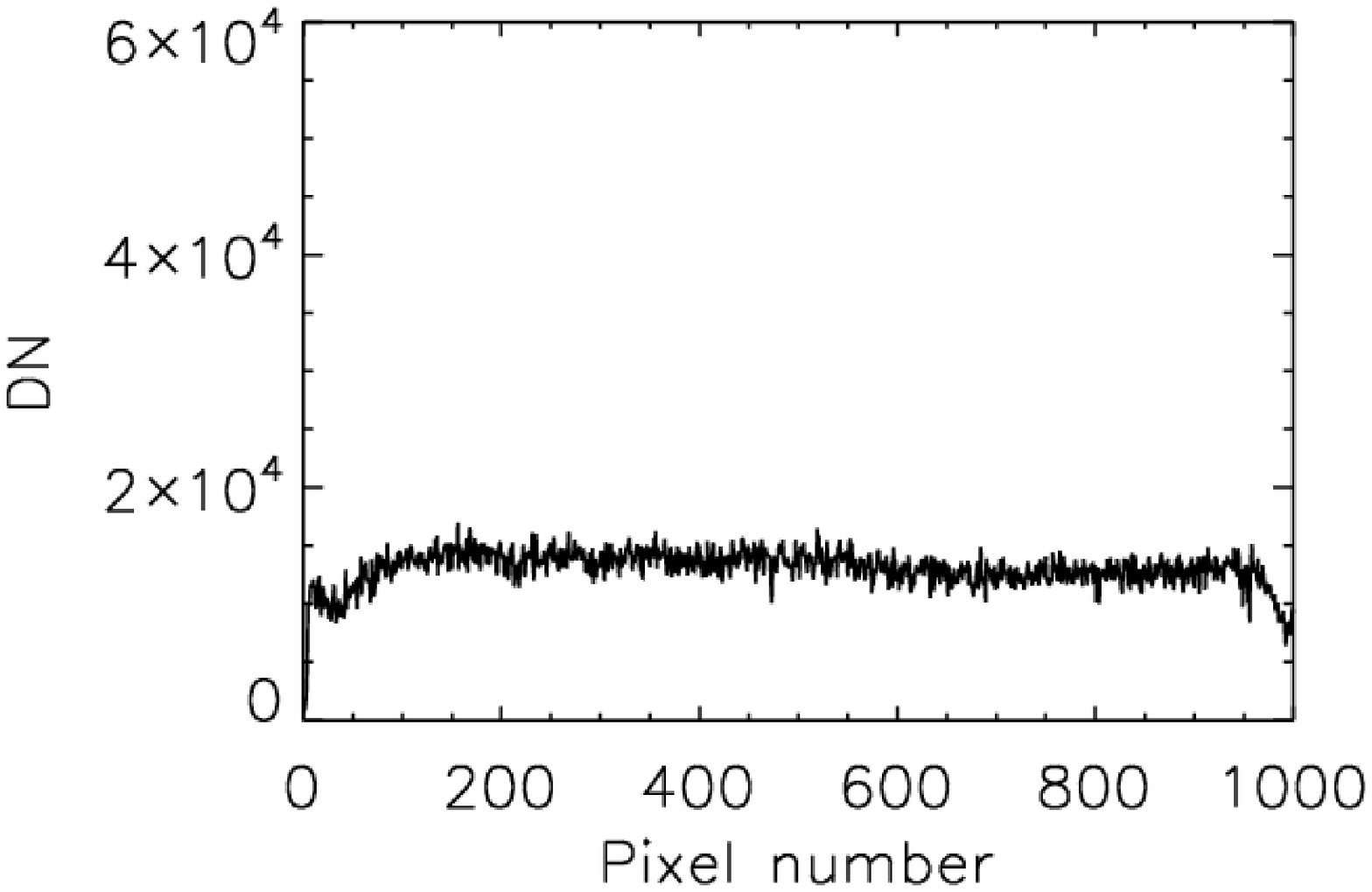} & \includegraphics[width=0.5\textwidth]{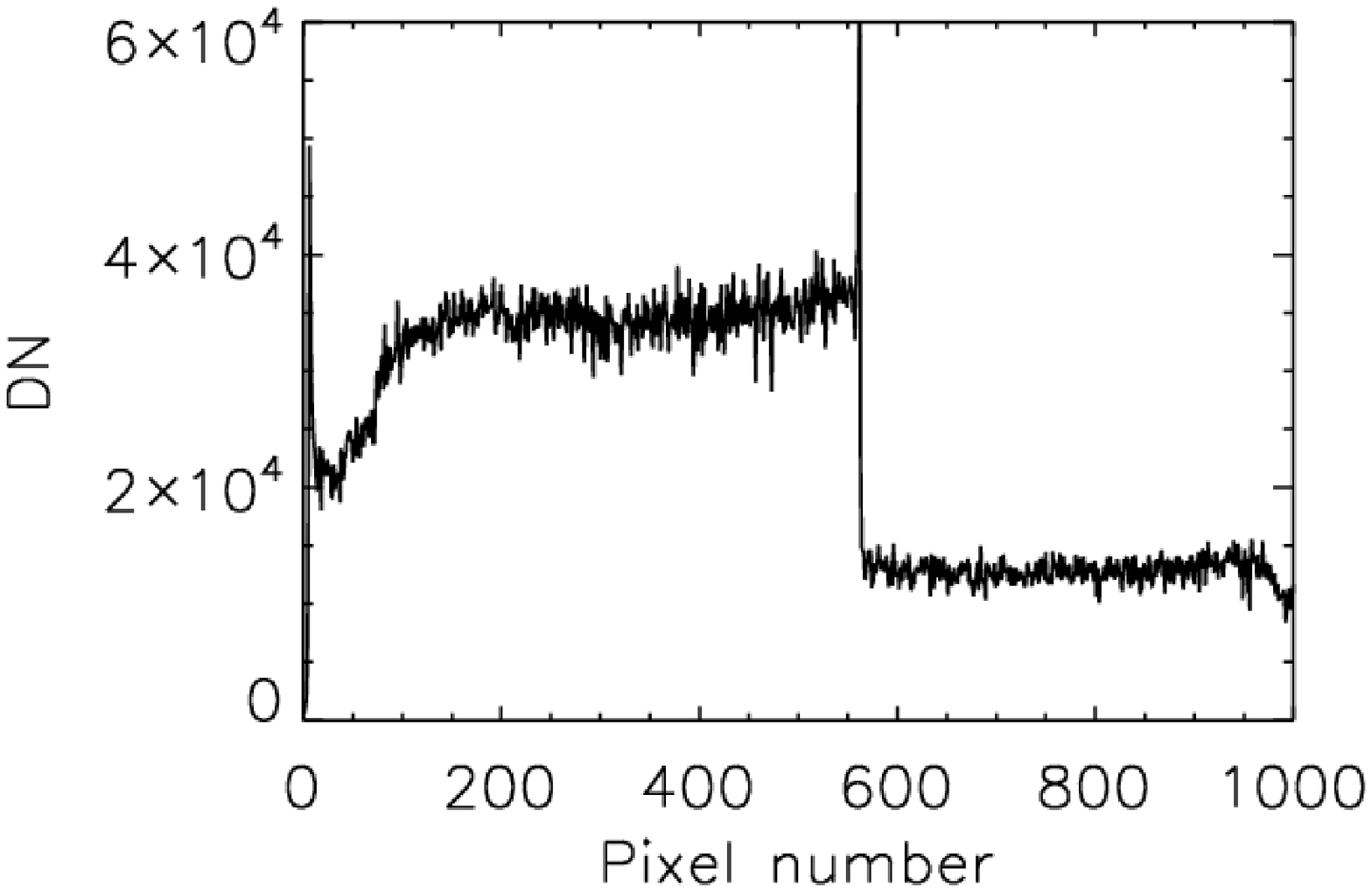} \\
\end{tabular}
\caption{\label{shift.fig}A vertical cut through a raw (pre- pseudo CDS subtraction) image before ({\it left}) and after ({\it right}) the permanent threshold shift damage occurred.  Note that the right half of the damaged array remained unaffected because this half of the detector was covered by the optical blocking filter.}
\end{figure}

After acquiring the large amount of $^{55}$Fe data presented in this work and while attempting to acquire additional X-ray line data, we unknowingly exposed the powered H1RG-125 to room light, causing a permanent threshold shift in the unfiltered half of the array.  The effect can be seen in the raw (pre-pseudo CDS subtraction) column data plotted in Figure \ref{shift.fig}.  These plots correspond to before and after the threshold shift damage occurred.  Note that most of the noise and structure seen in the background of these plots is fixed pattern noise.  These artifacts appear in all raw images and cleanly subtract out with the pseudo CDS.  In fact, the threshold shift is constant enough that it also subtracts out with the pseudo CDS, however the effect still decreases dynamic range and produces excess noise in the entire chip.  Prior to the threshold shift H1RG-125 exhibited a total noise of 12 DN RMS.  After the threshold shift, noise in the unaffected region did not change, but the affected region showed noise of 20 DN RMS.

It has been suggested that room temperature annealing may slowly fix the permanent threshold shift over time.  The array will be tested in a few months to evaluate whether or not any change is taking place.  Note that the half of the detector covered by the 500 \AA\ Al optical blocking filter remained unaffected because it was never exposed to optical light and did not experience the oversaturation condition.  Finally, we also note that this sensitivity to array saturation at high bias voltage is only present in ROICs that use source follower sense nodes.  Future detectors will be built with capacitive transimpedance amplifiers (CTIA), a technology that will eliminate both permanent threshold shift and IPC by holding the sense node at constant potential.

\section{CONCLUSIONS}
We are continuing to test and characterize Hybrid CMOS detectors and associated readout hardware to evaluate their readiness for future space based X-Ray missions.  The inherent advantages of HCDs (pileup reduction and radiation/micrometeoroid hardness) have already been integrated into the science goals of proposed missions now that the technology has been shown to meet readnoise and energy resolution requirements similar to those of CCDs.  In this work we have reported the readnoise of one HCD to be 7.48 e$^-$ and energy resolution to be 4.2 \%.  HCD technology still has great promise for further advancement, particularly in the area of removing interpixel capacitance with CTIA readout and drastically increasing effective frame rate with multi-channel, windowed readout.

\section*{ACKNOWLEDGMENTS}
This work was supported by NASA grant NNG05WC10G.  We would like to thank the engineers at Teledyne, in particular Raphael Ricardo, Michael Eads, Neil Songco, Jing Chen, and Richard Blank, for their invaluable help with troubleshooting the system and developing the assembly code for programming and reading the H1RG with the SIDECAR\texttrademark.  We would also like to thank Jim Beletic for valuable contributions while we were troubleshooting hardware.

\bibliography{library}
\bibliographystyle{spiebib}

\end{document}